\documentclass[english,aps,reprint,titlepage,longbibliography]{revtex4-1}   

\usepackage[T1]{fontenc}	
\usepackage[latin9]{inputenc}	
\usepackage{geometry}		
\usepackage{graphicx}
\usepackage[above,below]{placeins}	
\usepackage{times}
\usepackage{float}

\usepackage{hyperref}  
\hypersetup{colorlinks=true,urlcolor=blue,citecolor=blue,linkcolor=blue}   
\usepackage{enumitem}          
\setlist{nosep}                 

\usepackage{lineno}
\begin{document}

\begin{titlepage}

  \title{Using psychometric tools as a window into students' quantitative reasoning in introductory physics}

\author{Trevor I.\ Smith}
\affiliation{Department of Physics \& Astronomy and Department of STEAM Education, Rowan University, 201 Mullica Hill Rd., Glassboro, NJ 08028, USA}

\author{Philip Eaton}
\affiliation{Department of Physics, Montana State University, Bozeman, Montana 59717, USA}

\author{Suzanne White Brahmia}
\affiliation{Department of Physics, University of Washington, Box 351560, Seattle, WA 98195-1560, USA}

\author{Alexis Olsho}
\affiliation{Department of Physics, University of Washington, Box 351560, Seattle, WA 98195-1560, USA}

\author{Andrew Boudreaux}
\affiliation{Department of Physics and Astronomy, Western Washington University, 516 High St, Bellingham, WA 98225, USA}

\author{Chris DePalma}
\affiliation{Department of Physics \& Astronomy, Rowan University, 201 Mullica Hill Rd., Glassboro, NJ 08028, USA}

\author{Victor LaSasso}
\affiliation{Department of Physics \& Astronomy, Rowan University, 201 Mullica Hill Rd., Glassboro, NJ 08028, USA}

\author{Scott Straguzzi}
\affiliation{Department of Computer Science, Rowan University, 201 Mullica Hill Rd., Glassboro, NJ 08028, USA}

\author{Christopher Whitener}
\affiliation{Department of Physics \& Astronomy, Rowan University, 201 Mullica Hill Rd., Glassboro, NJ 08028, USA}


  \begin{abstract}
    The Physics Inventory of Quantitative Literacy (PIQL), a reasoning inventory under development, aims to assess students' physics quantitative literacy at the introductory level. The PIQL's design presents the challenge of isolating types of mathematical reasoning that are independent of each other in physics questions. In its current form, the PIQL spans three principle reasoning subdomains previously identified in mathematics and physics education research: ratios and proportions, covariation, and signed (negative) quantities. An important psychometric objective is to test the orthogonality of these three reasoning subdomains. We present results from exploratory factor analysis, confirmatory factor analysis, and module analysis that inform interpretations of the underlying structure of the PIQL from a student viewpoint, emphasizing ways in which these results agree and disagree with expert categorization. In addition to informing the development of existing and new PIQL assessment items, these results are also providing exciting insights into students' quantitative reasoning at the introductory level. \clearpage
  \end{abstract}

  \maketitle
\end{titlepage}

\section{Introduction}
One major goal of university-level physics courses is the development of mathematical reasoning skills, but despite decades of research on the complex interplay between physics conceptual understanding and mathematical reasoning \cite{Sherin2001, Thompson2010, boudreaux2015, brahmia2016a, Brahmia2017c, rebello2007}, measuring these skills has not gained as much popularity as strictly conceptual assessments \cite{Madsen2017, Madsen2019}.

Assessing conceptual understanding is inherently easier than assessing generalized mathematical reasoning. The former is tied to specific physics contexts taught over a finite period, while the latter is ubiquitous across contexts and time. Physics education researchers have conducted qualitative case studies to probe students' mathematical reasoning and their transitions to expert-like reasoning (c.f., Refs.\ \cite{Hayes2010,Hu2013,Smith2013}). While this method provides a rich view into how specific students reason in a particular context, little has been published that characterizes the process of emerging expert-like mathematical reasoning across multiple topics for large groups of students. We propose module analysis of a multiple-response mathematical reasoning instrument in physics contexts as a potential probe to help fill this gap. 

We have developed the Physics Inventory of Quantitative Literacy (PIQL) to meet the need for a robust and easily administered multiple-choice assessment to measure students' mathematical reasoning in the context of physics (a.k.a., physics quantitative literacy, PQL) \cite{Olsho2019b}.  The PIQL  is intended to test three key components of PQL: ratio and proportion \cite{Cohen2005}, covariation \cite{carlson2010}, and negativity \cite{vlassis2004, brahmia2017a, White2018, White2019a, White2019b}. We use confirmatory factor analysis in the current study to determine whether students' responses to PIQL questions form patterns that are consistent with the intended groupings \cite{Eaton2018}. 

We consider PQL to be a conceptual blend between physics concepts and mathematical reasoning \cite{Fauconnier2002}. In order to measure the complexity of ideas that students bring from both of these input spaces, we have chosen to include some multiple-response (MR) questions in which students are instructed to ``select all statements that \textbf{must be true}'' from a given list, and to ``\textbf{\textit{choose all that apply}}'' (emphasis in the original text). The MR question format has the potential to reveal more information about students' thinking than standard single-response (SR) questions, but it also poses problems with data analysis, as typical analyses of multiple-choice tests assume SR questions. In this paper we use two different methods to identify groups of questions evident in students' responses to PIQL questions and compare them to the groups defined by our three PQL constructs: factor analysis, which treats all questions the same way, assuming one correct response (or combination of responses) \cite{Eaton2018}, and module analysis, which considers each correct response individually, allowing for more detailed analyses of MR questions \cite{Brewe2016}. Each method has affordances and constraints, and comparing their results allows us to gain more insight into the ways in which students' response patterns do and do not mirror our PQL constructs.

Multiple-response questions have been used to measure the richness of reasoning associated with the use of mathematics in Junior-level electricity and magnetism \cite{Wilcox2014}. Module analysis helps identify patterns in student responses without constraining the responses to a dichotomous ``all right''-or-wrong scoring scheme that is inherent in factor analysis \cite{Brewe2016, Eaton2018}.

Data were collected in two different terms at a comprehensive public university in the Northwestern United States.  The PIQL was given as a pretest during the first week of the term in three different calculus-based introductory physics classes: Mechanics ($N=821$), Electricity and Magnetism ($N=701$), and Thermodynamics and Waves ($N=585$), each of which had an approximately 80\% participation rate. These data do not form a matched set, but we take them as three snapshots in time, which may be representative of a progression through the introductory course sequence: before mechanics (PreMech), after mechanics and before E\&M (PostMech), and after E\&M (PostEM). Overall scores on the PIQL increase over time , so we can consider students progressing toward expertise through our data (mean scores are 50\%, 56\%, and 58\%, respectively, with standard deviations around 17\%--18\% for each data set) \cite{Smith2018,Smith2019c}. For these preliminary analyses we choose to ignore potential biases due to some students not completing all three introductory courses (due to personal choice or majoring in a field that only requires one or two courses).
Future work will look at a matched set of students as they progress through the introductory sequence.


In the following sections we present the results of factor analysis and module analysis with regard to these three data sets. We compare the factors and modules with the groups of questions intended to test our three primary constructs of PQL: ratio and proportion (Q1--Q6), covariation (Q7--Q14), and negativity (Q15--Q20). We show how students' responses are and are not consistent with these groups and suggest future avenues for analyzing PIQL data.

\section{Factor Analysis}
Exploratory factory analysis (EFA) uses correlations between questions generated from student responses to identify questions that load onto the same factor \cite{hayton2004factor}. 
Confirmatory factor analysis (CFA) takes a researcher-specified factor model and estimates the question loading values to best reproduce the target correlation matrix. Using the model's estimated parameters, goodness-of-fit statistics can be calculated to verify whether or not the proposed model adequately represents the correlational grouping of the question on the assessment \cite{brown2014confirmatory}. 
Initially we compared the proposed question groups with the data by applying CFA and using the Confirmatory Fit Index (CFI) and Tucker-Lewis Index (TLI) as indicators of how well the data fit the model (with values above 0.90 indicating good agreement) \cite{Eaton2018}. For each term, the CFI and TLI were below 0.8, indicating that the data do not support the proposed three groups.

Lacking strong CFA results, we used EFA with an oblimin rotation to identify factors that may be determined directly from the data. We combined data from both terms, but separated them according to the course sequence: PreMech, PostMech, and PostEM. 
Parallel analysis suggested three factors be used in EFA for each of the classes, and the subsequent results of the EFA can be found in Table \ref{tab:loadings}.

Questions with loading values of at least 0.25 were considered to belong to that factor. Table \ref{tab:factors} shows the resulting factor structure for each of the three samples. While some questions seem to group together (e.g., 11 and 12 strongly load onto Factor 3 in all three data sets), the resulting factor structure is not consistent over time. Moreover, all factors contain questions that correspond with each of our PQL constructs. These results emphasize the dynamic nature of students' mathematical reasoning skills and suggests that the elements of PQL may be related in complex and subtle ways. 

\begin{table}[]
	\caption{Factor loadings by data set: questions with a loading of at least 0.25 are considered to load onto that factor. Horizontal lines show the breaks between groups corresponding with PQL constructs. The factor loadings do not support these groups.} 
	\begin{ruledtabular}
		\begin{tabular}{l|lll@{\hspace{2mm}}|@{\hspace{2mm}}lll@{\hspace{2mm}}|@{\hspace{2mm}}lll}
			 \multicolumn{1}{c}{~}      &\multicolumn{3}{c}{PreMech}&\multicolumn{3}{c}{PostMech}&\multicolumn{3}{c}{PostEM}\\
			       & F1     & F2      & F3 & F1 & F2 & F3 & F1 & F2 & F3\\ \hline
			Q1 & 0.16 & 0.40 &       & 0.66 &  &  & 0.20 & 0.27 & \\
			Q2 & 0.37 & 0.30 &        & 0.29 &  & 0.32 & 0.61 & -0.12 & -0.14\\
			Q3 & 0.59 & 0.22 &       & 0.33 & 0.32 & 0.37 & 0.58 & 0.34 & \\
			Q4 & 0.37 & 0.43 &  & 0.28 &  & 0.21 & 0.56 &  & 0.33\\
			Q5 & 0.48 & -0.31 &  & 0.16 & 0.58 &  & 0.43 & 0.23 & \\ 
			Q6 & 0.46 & 0.29 & 0.29 & 0.47 & 0.24 & 0.22 & 0.41 & 0.17 & 0.20\\ \hline
			Q7 & 0.54 &  &  & 0.27 & 0.15 & 0.51 & 0.20 & 0.36 & \\
			Q8 & 0.59 &  & 0.17 &  & 0.53 & 0.32 & 0.26 & 0.40 & 0.30\\
			Q9 & 0.32 & 0.51 &  & 0.47 & 0.39 &  & 0.37 & 0.12 & 0.29\\
			Q10 & 0.53 & 0.21 & 0.26 & 0.40 & 0.56 & 0.23 & 0.53 & 0.36 & \\ 
			Q11 & 0.22 & 0.10 & 0.89 &  & 0.17 & 0.86 & 0.16 & 0.18 & 0.83\\
			Q12 & 0.14 & 0.21 & 0.86 &  & 0.15 & 0.85 & 0.14 & 0.17 & 0.85\\
			Q13 & 0.41 & 0.22 &  & 0.15 & 0.45 & 0.21 & 0.56 & 0.10 & \\
			Q14 & 0.56 &  & 0.13 & 0.11 & 0.53 &  & 0.45 & 0.15 & 0.18\\ \hline
			Q15 & 0.34 & 0.37 &  & 0.60 & 0.22 &  & 0.35 & 0.18 & 0.28\\ 
			Q16 & -0.12 & 0.34 & 0.19 & 0.49 & 0.21 &  &  &  & 0.46\\
			Q17 &  & 0.61 & 0.28 & 0.34 & 0.44 &  & 0.46 & 0.13 & 0.20\\
			Q18 & 0.37 & 0.32 & 0.28 & 0.56 & 0.20 & 0.29 & 0.21 & 0.64 & 0.34\\
			Q19 & 0.27 & 0.19 & 0.19 & 0.44 &          & 0.14 & 0.17  & 0.65 & \\
			Q20 &        & 0.52  &           & -0.13 & 0.58 &         & -0.16 & 0.70 & 
		\end{tabular}
		 \end{ruledtabular}
	\label{tab:loadings}
\end{table}
\begin{table}[tb]
    \caption{Factors by data set: identified from factor loadings (Table \ref{tab:loadings}). Questions that overlap with other factors are shown in parentheses.}
    \begin{ruledtabular}
    \begin{tabular}{c p{0.25\columnwidth} p{0.25\columnwidth} p{0.25\columnwidth}}
       Factor& PreMech & PostMech & PostEM\\
        \hline
        1 & 2, 3, (4), 5, 6, 7, 8, 10, 13, 14, 15, 18, 19& 1, (2), (3), 4, 6, 9, (10), 15, 16, (17), 18, 19 & 2, 3, 4, 5, 6, (8), 9, 10, 13, 14, 15, 17\\[1ex]
        2 & 1, (2), 4, (6), 9, 15, 16, 17, (18), 20 & (3), 5, 8, (9), 10, 13, 14, 17, 20 & 1, (3), 7, 8, (10), 18, 19, 20\\[1ex]
        3 & (6), (10), 11, 12, (17), (18) & 2, (3), 7, (8), 11, 12, (18) & (4), (8), (9), 11, 12, 16, (18)\\
    \end{tabular}
    \end{ruledtabular}
    \label{tab:factors}
\end{table}

\section{Module Analysis}
A limitation of factor analysis applied to multiple-choice assessments is that it requires dichotomous data in which every response is coded as either correct or incorrect. For multiple-response (MR) questions, dichotomous scoring methods require a student to choose all correct responses and only correct responses to be considered correct. This ignores the nuance and complexity of students' response patterns within (and between) questions. In an effort to account for the complexity of student responses to MR questions, we applied module analysis for multiple-choice responses (MAMCR) to examine the network of student responses to PIQL questions \cite{Brewe2016}. MAMCR uses community detection algorithms to identify modules (a.k.a.\ communities, clusters, etc.) within networks of responses to multiple-choice questions. 
Brewe, Bruun, and Bearden used MAMCR to identify modules of incorrect responses to questions on the Force Concept Inventory \cite{Brewe2016}; we have chosen to analyze a network of only correct responses to PIQL questions. The benefit of this method is that we can examine the patterns that arise from students' selections of each individual correct response, which preserves some of the complexity of MR questions: a student who chooses one correct response to a question with two correct responses is treated differently than a student who chooses only incorrect responses. A limitation of the way that we are using MAMCR is that we are ignoring whether or not a student chooses incorrect responses in addition to correct responses. Expanding the network to include correct and incorrect responses could address this limitation, but this is beyond the scope of the current study.

\begin{figure*}[tb]
    \centering
    \includegraphics[width = 0.5\textwidth]{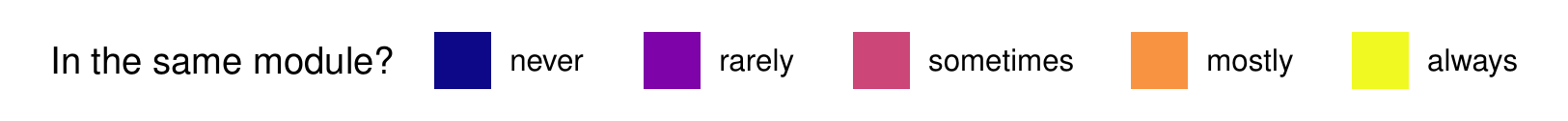}
    
    \includegraphics[width = 0.32\textwidth]{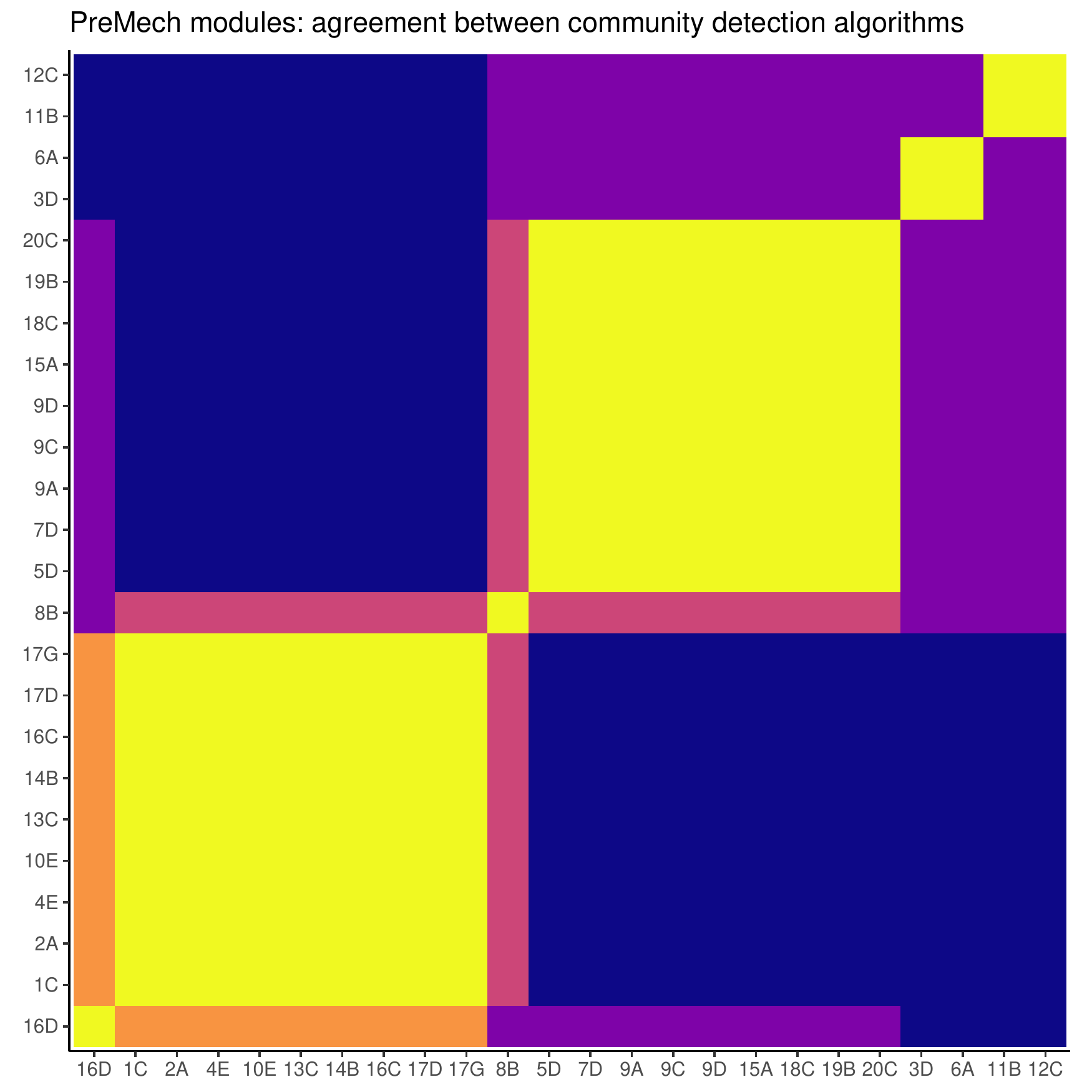}\hspace{3mm}\includegraphics[width = 0.32\textwidth]{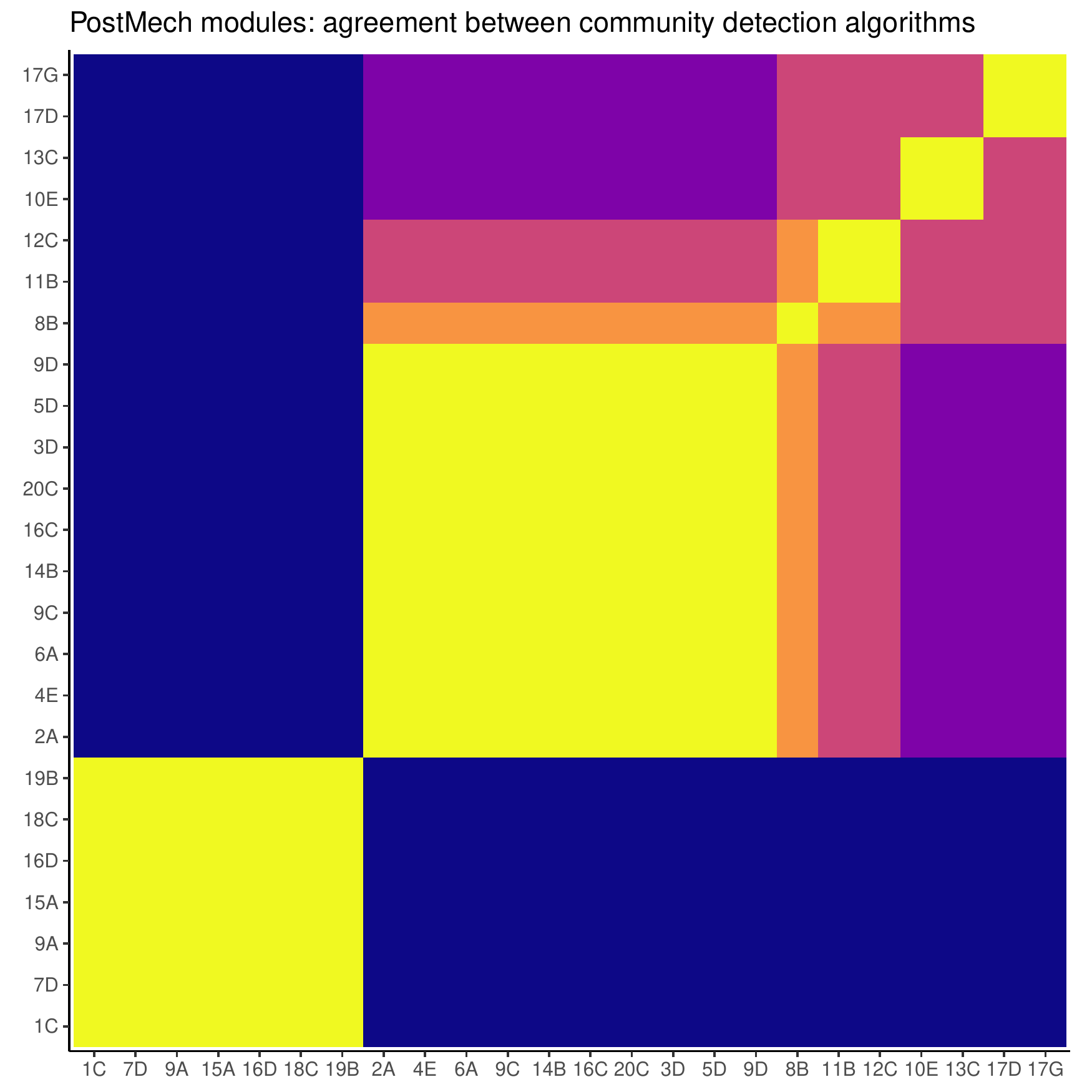}\hspace{3mm}\includegraphics[width = 0.32\textwidth]{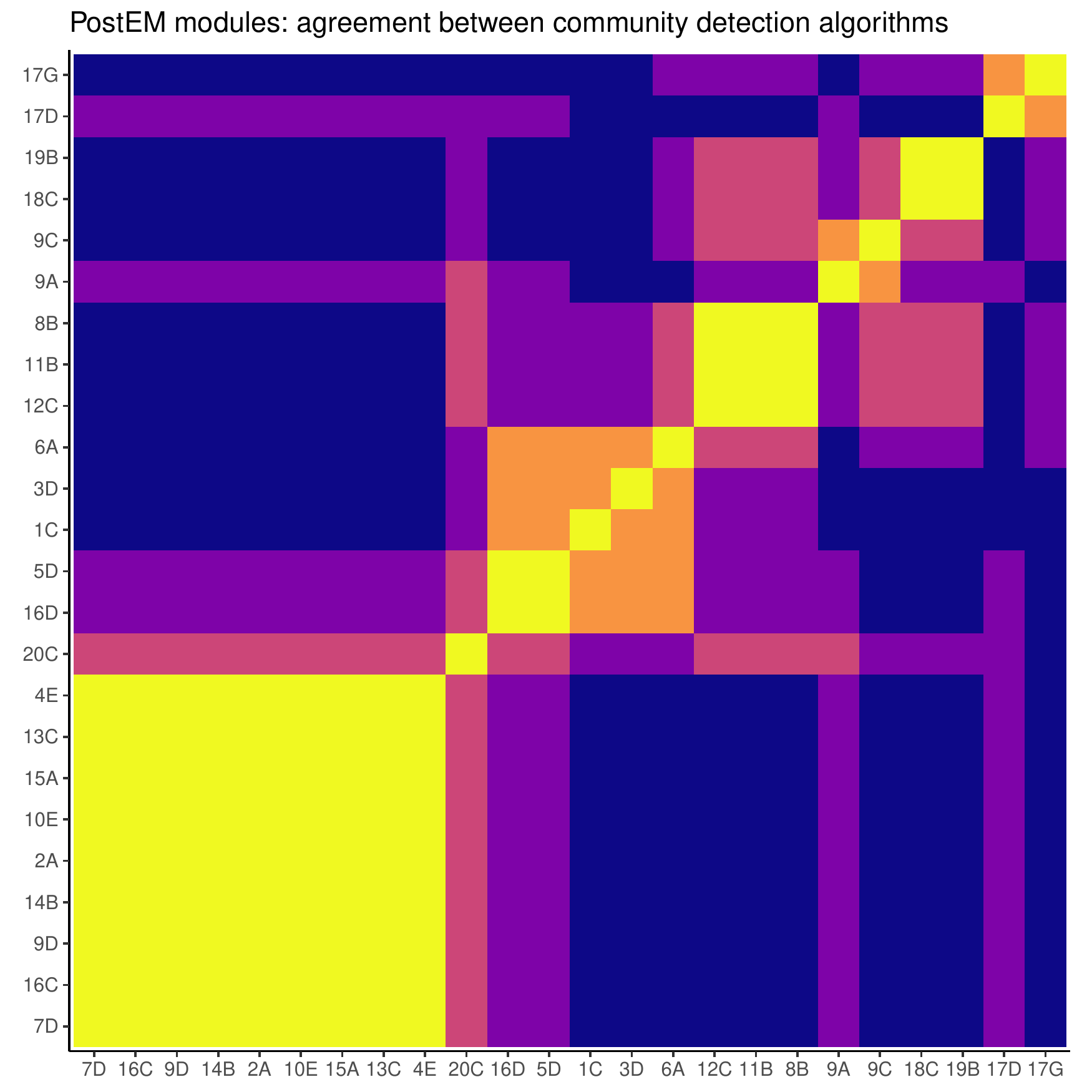}
    
    \hspace{2mm}(a)\hspace{0.3\textwidth}\hspace{3mm}(b)\hspace{3mm}\hspace{0.3\textwidth}(c)
    \caption{Heatmaps showing the level of agreement between community detection algorithms for each data set: (a) PreMech, (b) PostMech, and (c) PostEM. Yellow indicates that two responses were in the same module for all algorithms; dark blue indicates that two responses were never in the same module.}
    \label{fig:heatmaps}
\end{figure*}

We attempted to follow the methods for MAMCR presented by Brewe, Bruun, and Bearden as closely as possible \cite{Brewe2016}: we used the igraph package in the R programming language to create networks from students' response patterns \cite{igraph, r}; we created a ``backbone'' network by applying the LANS algorithm (with $\alpha = 0.05$) using code developed by Traxler, Gavrin, and Lindell \cite{Foti2011, Traxler2018}; and we attempted to use the InfoMap community detection algorithm multiple times to identify modules that are consistent across random fluctuations in the analysis \cite{infomap, Rosvall2009}. Unfortunately, like others, we were not able to obtain meaningful results using InfoMap, either within the igraph package or as a standalone program \cite{Wells2019}. In the absence of clear guidelines regarding which community detection algorithm would be most relevant, we chose to compare the modules identified by six different algorithms available in igraph: Louvain, InfoMap, Spinglass, Fast and Greedy, Leading Eigenvalue, and Label Propagation. 
We feel confident that modules that are identified by multiple community detection algorithms are representative of the data. 


Figure \ref{fig:heatmaps} shows heatmaps representing the co-occurrence matrix for each data set. Each heatmap is symmetric about the diagonal, with the color of each cell representing how often each pair of responses (designated by row and column) cluster in the same module. Bright yellow indicates that two responses occur in the same module for all algorithms, and dark blue indicates that two responses are never clustered together \footnote{Two of the community detection algorithms (InfoMap and Label Propagation) identified only one module for each data set, with all responses being grouped together; therefore, these algorithms are effectively removed from our analysis.}. The order of the responses along the axes of the heatmaps have been chosen to make it easier to visually identify modules; the order is not the same for all three data sets. We ran each algorithm 100 times to ensure that the modules were consistent. Results from the Spinglass algorithm showed slight variations, but provided consistent modules across 80--90\% of analyses. Results from all other algorithms showed no variation across 100 runs. 

\begin{table}[bt]
    \caption{Modules by data set: identified from heatmaps (Fig.\ \ref{fig:heatmaps}). Letters have been removed from labels for questions with only one correct response. Responses that overlap with other modules are shown with parentheses.}
    \begin{ruledtabular}
    \begin{tabular}{c p{0.25\columnwidth} p{0.25\columnwidth} p{0.25\columnwidth}}
       Module& PreMech & PostMech & PostEM\\
        \hline
        1 & 1, 2, 4, 10, 13, 14, 16C, 16D, 17D, 17G, (8) & 1, 7, 9A, 15, 16D, 18, 19 & 2, 4, 7, 9D, 10, 13, 14, 15, 16C, (20)\\[1ex]
        2 & 5, 7, 9A, 9C, 9D, 15, 18, 19, 20, (8) & 2, 3, 4, 5, 6, 9C, 9D, 14, 16C, 20, (8) & 1, 3, 5, 6, 16D, (20)\\[1ex]
        3 & 3, 6 & (8), 11, 12 & 9A, 9C, (20)\\[1ex]
        4 & 11, 12 & 10, 13 & 8, 11, 12, (20)\\[1ex]
        5 & & 17D, 17G & 18, 19, (20)\\[1ex]
        6 & &  & 17D, 17G
    \end{tabular}
    \end{ruledtabular}
    \label{tab:modules}
\end{table}

Table \ref{tab:modules} shows the modules that we identified by examining these heatmaps. The modules are not the same in the three different data sets, suggesting that students' knowledge and experience affect the ways in which they answer PIQL questions. The PreMech results show two large modules (big yellow squares in Fig.\ \ref{fig:heatmaps}a) that are almost completely separate from one another, with two pairs of responses clustering with one of these bigger modules for one algorithm. Response 8B has an equal likelihood of being in either module. As with the EFA results, both of the large modules in PreMech contain responses associated with all three of the PQL constructs.

The PostMech (Fig.\ \ref{fig:heatmaps}b) and PostEM (Fig.\ \ref{fig:heatmaps}c) results also show 1--2 large modules that are mostly isolated from other responses, and a few smaller modules, but the differences between the heatmaps provide interesting insights into how student mathematical reasoning changes over time. For example, the large module in the PostEM results contains responses from both of the large modules in PreMech and both of the large modules in PostMech. Additionally, the number of small modules increases from PreMech to PostMech to PostEM. Contrast this to what we expect from a hypothetical group of experts:  all questions would be answered correctly, resulting in all responses being in one coherent module. Our data show that as students progress toward expertise during the introductory sequence, modules become less coherent, not more. Additional data from upper-division students are needed to examine the continuation of this progression.


The MR questions with more than one correct response show some particularly interesting trends. Question 9 has three correct responses (9A, 9C, 9D) that group differently in each data set. 
Question 16 has two correct responses (16C, 16D) that appear in different modules in both the PostMech and PostEM results, and question 17 has two correct responses (17D, 17G) that almost always group together (and often separate from other responses).

\begin{figure}[tb]
    \centering
    \includegraphics[width = 0.67\columnwidth]{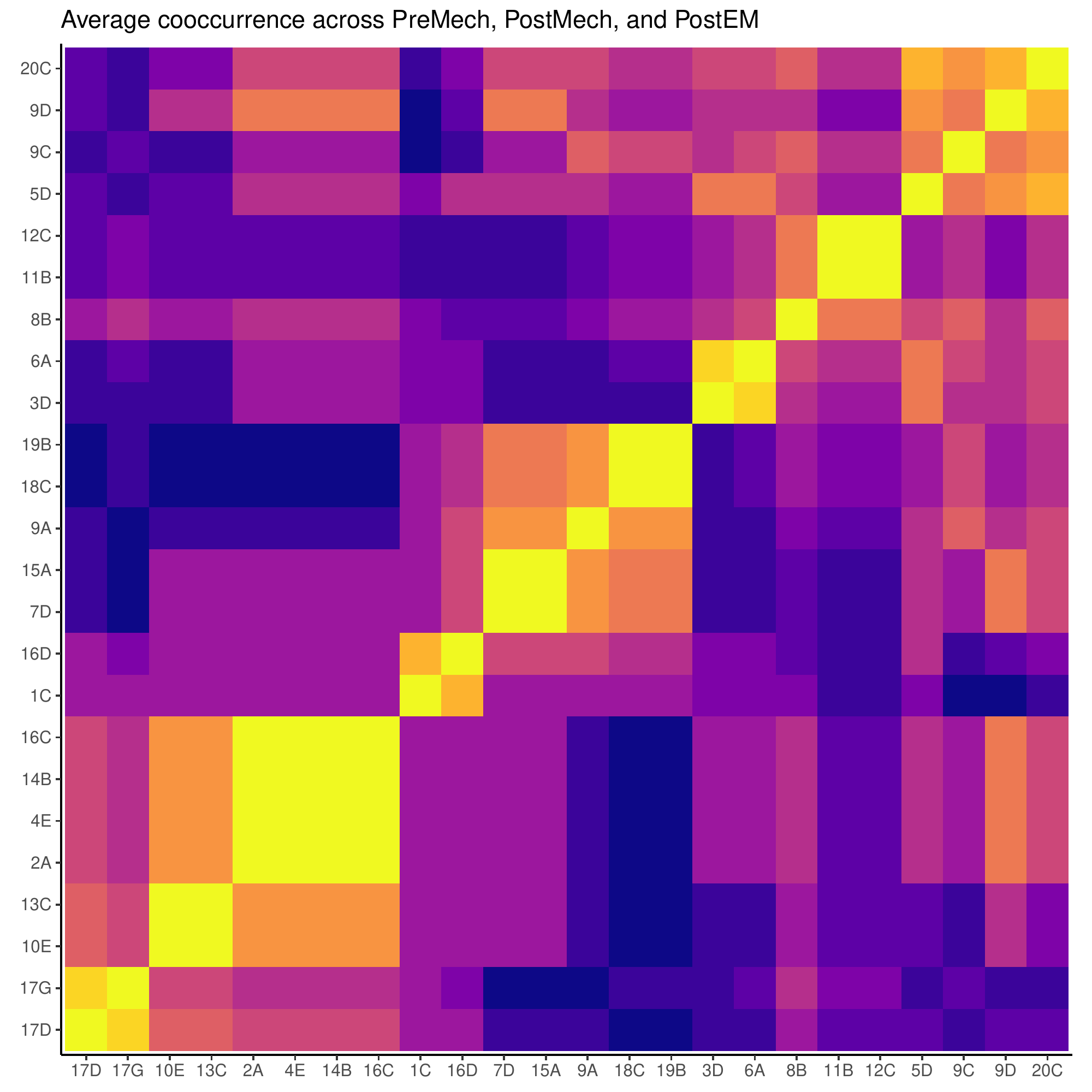}
    \caption{Combined heatmap showing the average cooccurrence matrix. The bright squares along the diagonal are used to identify the submodules listed in Table \ref{tab:submod}. The response order along the axes was determined by a hierarchical clustering method using the heatmap function in the stats package in R \cite{r}.}
    \label{fig:submod}
\end{figure}

The changes in module definitions over time led us to look for consistent patterns across the results, which may represent stable elements of student reasoning. For example: questions 11 and 12 always appear in the same module---often with question 8, and separate from others. 
Figure \ref{fig:submod} combines all three heatmaps from Fig.\ \ref{fig:heatmaps} to show the average likelihood that each question pair occurs in the same module. Table \ref{tab:submod} lists the ``submodules'' that we have identified as being consistent across our analyses. Each of these submodules may be seen as a bright yellow/orange square along the diagonal in Fig.\ \ref{fig:submod}, with submodule D (in the upper right corner) being the least cohesive (least bright). 
Some submodules are subsets of our PQL constructs: ratio and proportion (C), covariation (F and G), and negativity (H). Others include questions from two or three of these constructs (A, B, D, E, and I), emphasizing the connections between these constructs. 

\section{Comparing Results}
{Comparing Tables \ref{tab:factors} and \ref{tab:modules} reveals substantial differences between the EFA and MAMCR results. Even the submodules in Table \ref{tab:submod} do not show up consistently in the factor definitions: for example, questions 5, 9, and 20 from submodule B do not have their strongest loadings on the same factor in any of the data sets. However, we can see some similar results. Questions 11 and 12 are always in the same factor (with very strong loadings in Table \ref{tab:loadings}). Questions 18 and 19 are always in the same factor. Questions 10 and 13 always have their \parfillskip 0pt

}
\begin{table}[H]
    \caption{Consistent ``submodules'' of question/response nodes identified across data sets. Questions intended to probe ratio and proportion are shown in bold, questions for covariation are in italics, and questions for negativity are underlined. Response 9A fits equally with two submodules.}
    \begin{ruledtabular}
    \begin{tabular}{c l c l} 
        Submodule & Questions & Submodule & Questions\\
        \hline
        A & \textbf{1}, \underline{16D} & F & \textit{8}, \textit{11}, \textit{12}\\
        B & \textbf{2}, \textbf{4}, \textit{14}, \underline{16C} & G & \textit{10}, \textit{13}\\
        C & \textbf{3}, \textbf{6} & H & \underline{17D}, \underline{17G} \\
        D & \textbf{5}, \textit{9C}, \textit{9D}, \underline{20} & I & \underline{18}, \underline{19}, (\textit{9A})\\
        E & \textit{7}, \underline{15}, (\textit{9A}) 
    \end{tabular}
    \end{ruledtabular}
    \label{tab:submod}
\end{table}

\noindent strongest loadings in the same factor (although 10 could load on more than one factor for the PostMech and PostEM results). Perhaps the most consistent result across both analyses is that the clusters of questions/responses identified from the data change over time. Moreover, these clusters do not correspond with the groups defined by our PQL constructs. 
This suggests that either a) students' PQL cannot be separated into skills regarding ratio and proportion, covariation, and negativity, or b) their skills in these areas have developed similarly such that they are functionally equivalent. Regardless of the interpretation, MAMCR reveals complexity and structure that changes over time as students progress through the introductory sequence.

\section{Summary and Future Directions}
We have shown that two methods for identifying clusters of questions based on students' responses can yield differing results. The dynamics of these clusters provide windows into students' progression toward expertise. 
MAMCR also opens the possibilities for future work that goes beyond analysis of only correct responses by identifying modules of incorrect responses as well \cite{Brewe2016}. We plan to look more closely at the dynamics of these modules over time by using matched sets of responses collected from the same students at different times, and by expanding data collection beyond the introductory sequence. These longitudinal data will allow us greater confidence in claims regarding how students' response patterns change over time. The coupling of PIQL MR questions with MAMCR shows promise for finding patterns of emergent expertise in mathematical reasoning in introductory physics, and beyond, on a scale that cannot be achieved using qualitative research methods.


\begin{acknowledgments}
This work is supported by the National Science Foundation under grants 1832836, 1832880, and 1833050.
\end{acknowledgments}

\bibliography{TIS.bib} 

\end{document}